\documentclass[a4paper,11pt]{article}
\usepackage{pos}

\usepackage[left]{lineno}

\title{Deep Sets and event-level maximum-likelihood estimation for fast pile-up jet rejection at the
ATLAS High Level Trigger}
\ShortTitle{DIPz and MLPL for fast pile-up jet rejection in ATLAS}

\author*{Mohammed Aboelela}

\onbehalf{on behalf of the ATLAS Collaboration}

\affiliation[]{Southern Methodist University,\\
  Dallas, TX, USA}

\emailAdd{mo.aboelela@cern.ch}

\abstract{Multiple proton--proton interactions occur in the same bunch crossing at the LHC (“pile-up”),
with the mean number of interactions per bunch crossing reaching up to about 60 in Run 3 and up to
about 200 at the High-Luminosity LHC. As a direct consequence, hadronic activity within each bunch crossing
increases, making it more difficult to identify collisions of interest pertaining to multijet final states.
This work introduces DIPz, an uncertainty-aware Deep Sets–based machine-learning model that estimates each
jet’s origin along the beam axis as a Gaussian likelihood using the set of charged-particle tracks matched
to the jet. Next, the DIPz predictions for each jet are combined into a likelihood-based event-level
discriminant, which is optimised to select events compatible with a single collision producing
a given jet multiplicity. This combined approach provides a robust and computationally efficient
method for pile-up rejection in multijet final states, applicable to real-time event selections
at the ATLAS High Level Trigger (HLT). Deployed online during Run~3 data taking, this approach
reduces background rates by more than a factor of two, and decreases CPU consumption by approximately
one third relative to the baseline data-taking configuration. It provides an early, flavour-agnostic
pile-up jet rejection layer at the HLT preselection stage, introducing a previously unavailable handle
for high-multiplicity jet signatures targeted by ATLAS.}

\FullConference{
The 43rd International Conference on High Energy Physics (ICHEP 2026)\\
30 July – 05 August 2026\\
Natal, Brazil
}

\begin{document}
\maketitle

\section{Introduction}

At the Large Hadron Collider (LHC), proton bunches collide at a rate of up to 40~MHz,
producing multiple interactions per bunch crossing, known as pile-up. During Run 3, the average
number of interactions per crossing reaches up to 60, and is expected to reach values near 200
at the High-Luminosity LHC (HL-LHC). Each one of these overlapping interactions can produce its
own jets, leading to dense event topologies that complicate the reconstruction of multijet final
states and challenge real-time event selection at the trigger level.

The ATLAS experiment~\cite{ATLASdetector} uses a two-level trigger system to reduce the event rate
from 40~MHz to about 3~kHz for permanent storage. The second level is the software-based High Level
Trigger (HLT), where refined reconstruction algorithms are executed within strict latency constraints.
At the HLT, charged-particle track reconstruction takes place, which in turn provides essential
information for jet and vertex reconstruction. Performing track finding over the entire inner detector
(full-scan tracking) is by far the most CPU-intensive algorithm that runs at that stage~\cite{Aad_2024}.
To mitigate this cost, a preselection stage is introduced, in which regional tracking is performed
in localised super-Regions of Interest (super-RoIs)~\cite{TRIG-2018-08} around calorimeter jets. Used 
primarily for fast $b$-tagging~\cite{FastBtag}, super-RoI tracks can also be exploited to identify and 
reject pile-up jets early in the trigger selection, improving the efficiency of identifying true multijet 
signal events.

This work, documented in more detail in Ref.~\cite{ATL-DAQ-PUB-2026-002}, presents a novel approach for 
fast and efficient pile-up jet rejection at the ATLAS HLT. An uncertainty-aware Deep Sets–based neural 
network, named \textbf{DIPz}, regresses the longitudinal (along the beamline $z$-axis) production vertex 
location of each jet from the charged-particle tracks reconstructed within the corresponding jet super-RoI. 
Per-jet likelihoods from DIPz are subsequently combined into an event-level discriminant, the 
\textbf{Maximum Log Product of Likelihoods} (\textbf{MLPL}), designed to efficiently reject events 
dominated by pile-up jets. Although broadly applicable, this approach is particularly well suited for 
optimising the HLT capabilities for high jet-multiplicity physics signatures. It provides a fast rejection 
mechanism analogous to fast $b$-tagging strategies, but designed explicitly for flavour-agnostic 
multijet signatures, addressing an important gap in the current trigger preselection options.

\section{DIPz for uncertainty-aware jet origin regression}

Rather than relying on a simple $z$-average of track positions within the jet, which provides insufficient
vertex resolution, DIPz is designed to exploit the full set of track-level properties and produce
a per-jet uncertainty estimate alongside each prediction. DIPz is applied to calorimeter jets 
that survive the selection criteria of having transverse momentum $p_{\mathrm{T}}>20$~GeV and 
pseudorapidity 
$|\eta|<2.5$.\footnote{In the ATLAS coordinate system, the pseudorapidity is 
defined as $\eta = - \ln \tan(\theta/2)$, where $\theta$ is the polar angle measured 
between the beamline $z$-axis and the particle's momentum vector.} 
Each track is described by its $p_{\mathrm{T}}$, its impact parameters $d_0$ and $z_0$ 
relative to the beamspot (the centre of the luminous region where $pp$ collisions occur), 
its separations $\Delta\eta$ and $\Delta\phi$ from the jet axis, $q/p$, and the $\chi^2$  and number 
of degrees of freedom of the track fit. The DIPz algorithm follows the Deep Sets 
framework~\cite{zaheer2018deepsets, ATL-PHYS-PUB-2020-014}, which is well suited for variable-length 
unordered inputs such as tracks within a jet. Each track is processed through shared feed-forward layers, 
producing latent representations that are summed to ensure permutation invariance. The jet $p_{\mathrm{T}}$ and $\eta$
are processed by a separate feed-forward network, whose output is concatenated with the summed track
representation and passed to a final network that predicts the mean ($\mu_z$) and standard deviation
($\sigma_z$) of a Gaussian describing the jet-vertex position along the beamline. The training target
corresponds to the true $z$-position of the collision vertex from which the jet originated. The loss
function used in training is the Gaussian negative log-likelihood,
\begin{equation}
\text{Loss} (z_\text{target}, \mu_z, \sigma_z)
=
-\log\left(\frac{1}{\sqrt{2\pi} \cdot \sigma_z} \cdot
\exp\left(-\frac{1}{2}\left(\frac{z_\text{target}-\mu_z}{\sigma_z}\right)^2\right)\right),
\label{eq:loss}
\end{equation}
allowing the model to learn both the central value and its associated uncertainty. The model has
19\,650 trainable weights. It is implemented in Keras~\cite{chollet2015keras} with a TensorFlow backend,
trained using the Adam optimiser~\cite{kingma2017adammethodstochasticoptimization} with early stopping,
and exported to the HLT software with lwtnn~\cite{lwtnn}.

The model is trained using simulated semileptonic $t\bar{t}$ events at $\sqrt{s}=13.6$~TeV, generated with
\textsc{Powheg}\allowbreak+\allowbreak\textsc{Pythia\,8}~\cite{6,7} (details in Ref.~\cite{ATL-DAQ-PUB-2026-002}), with pile-up overlaid
to emulate Run~3 conditions~\cite{8}. Training is performed on a dataset of 3 million jets,
with validation on a separate set of 600\,000 jets.

\begin{figure}[!ht]
  \centering
\includegraphics[width=0.69\textwidth]{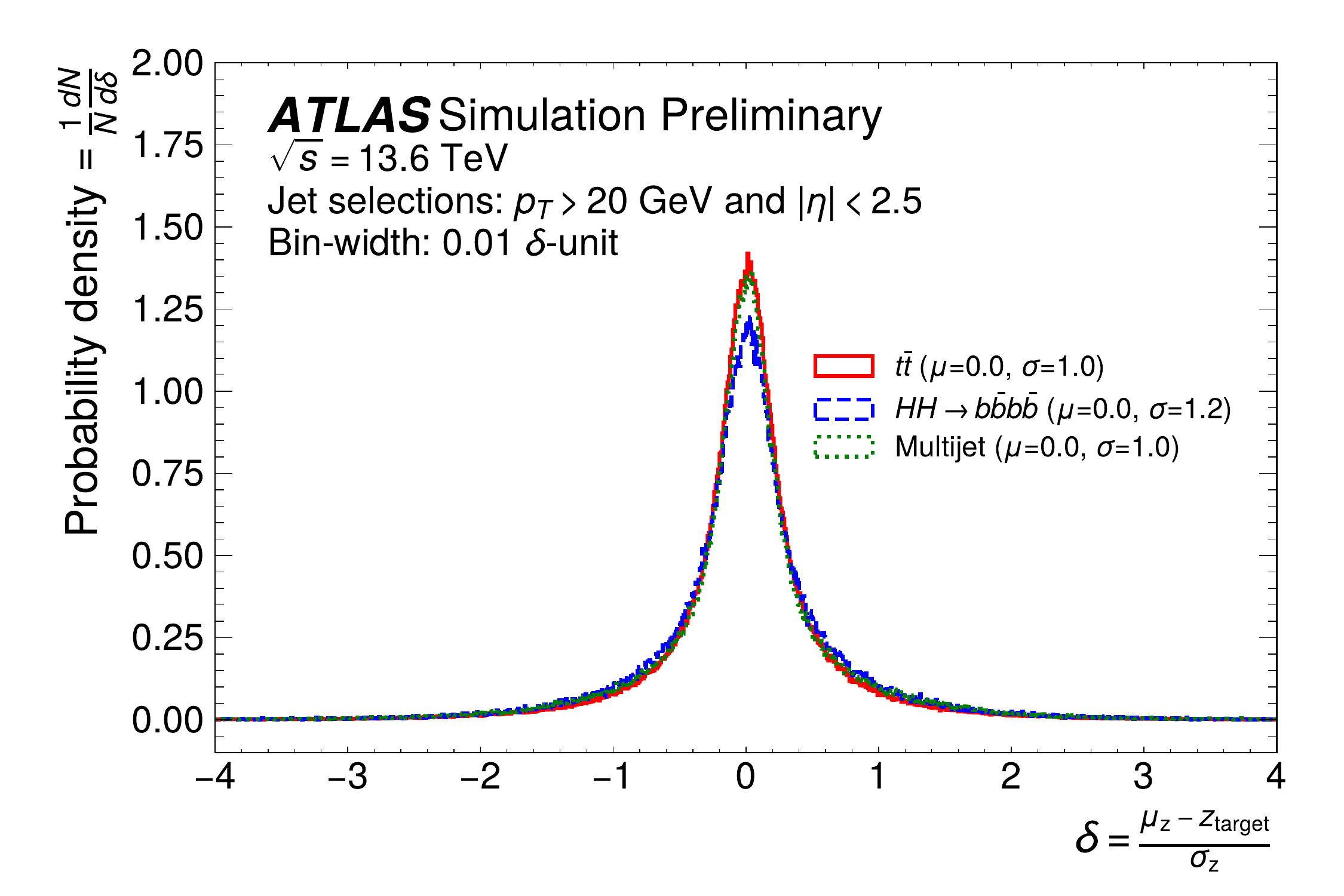}
  \caption{Normalised distributions of the significance of the prediction error $\delta$ across 
  multiple test jet samples. The $\mu$ and $\sigma$ in the legend denote the sample mean and standard deviation, respectively. Taken from Ref.~\cite{ATL-DAQ-PUB-2026-002}.}
  \label{fig:dipz_performance}
\end{figure}

The performance of DIPz is evaluated using jets originating from multiple physics processes:
a statistically independent $t\bar{t}$ sample, as well as multijet and $HH\to b\bar{b}b\bar{b}$ samples, all unseen during training.
Figure~\ref{fig:dipz_performance} shows the significance of the prediction error,
$\delta = (\mu_z - z_{\mathrm{target}})/\sigma_z$, which follows a standard normal distribution
for a perfectly calibrated model. The distributions are centred at zero, with standard deviations
of 1.0 ($t\bar{t}$), 1.2 ($HH\to b\bar{b}b\bar{b}$) and 1.0 (multijet). On average, the predicted uncertainties 
match the observed spread of the prediction errors, also for processes not used in the training. 
This shows that the model generalises  well to unseen physics processes, indicating that it has 
learned a robust representation of the underlying physics rather than overfitting to the training 
data. Moreover, Ref.~\cite{ATL-DAQ-PUB-2026-002} shows that this performance is mostly stable across 
jet $p_{\mathrm{T}}$, $\eta$, flavour, and the number of jet-associated tracks, as well as the event-level pile-up conditions, further demonstrating 
the robustness of the model.

\section{A fast rejection algorithm via Maximum Log Product of Likelihoods (MLPL)}
This approach relies on constructing an event-level discriminant based on the per-jet vertex predictions
provided by DIPz. For a given $n$-jet final state selection, a cut is optimised and applied to this
discriminant to reject pile-up-dominated events efficiently, enabling fast decision-making within the
tight latency constraints of the ATLAS trigger system.

The goal of this approach is to identify events in which at least $n$ jets originate from the same
$pp$ collision, using only the DIPz neural network outputs. For that purpose, each jet is
assigned a Gaussian likelihood function, characterised by its DIPz predicted mean $\mu_z$ and standard
deviation $\sigma_z$ along the beamline:
\begin{equation}
\mathscr{L}_j(z; \mu_{z_j}, \sigma_{z_j}) = \frac{1}{\sqrt{2\pi }\sigma_{z_j}}
\exp\left( -\frac{(z - \mu_{z_j})^2}{2\sigma_{z_j}^2} \right).
\end{equation}
To assess the compatibility of multiple jets originating from the same vertex within an event, the individual jet likelihoods
are multiplied to form a product likelihood for a group of $n$ jets $J_n$:
\begin{equation}
L(z; J_n) = \prod_{j \in J_n} \mathscr{L}_j(z; \mu_{z_j}, \sigma_{z_j}).
\label{eq:prod_like}
\end{equation}
The maximum of $L(z;J_n)$ can be shown to lie at the precision-weighted mean of the jet positions,
\begin{equation}
  \hat{z} = \frac{\sum_{j\in J_n}\mu_{z_j}/\sigma_{z_j}^2}{\sum_{j\in J_n}1/\sigma_{z_j}^2},
\end{equation}
so no numerical optimisation is needed. $L(\hat{z};J_n)$ is largest when the jets in $J_n$ are
precisely measured and compatible with a common origin. For an event with $p$ selected jets,
the groups $J_n$ are drawn from $J_m$, the $m$ highest-$p_{\mathrm{T}}$ jets, giving
$\binom{\min(m,p)}{n}$ combinations. The discriminant is the largest value among them:
\begin{equation}
  \mathrm{MLPL}(n,m) = \max_{J_n\subseteq J_m}\,\log L(\hat{z};J_n).
\end{equation}
The inclusion of the parameter $m$ controls the computational complexity, which otherwise grows
combinatorially with the number of jets. By appropriately choosing $n$ and $m$, a balance
between performance and computational efficiency is achieved.

\begin{figure}[!ht]
    \centering
    \includegraphics[width=0.69\linewidth]{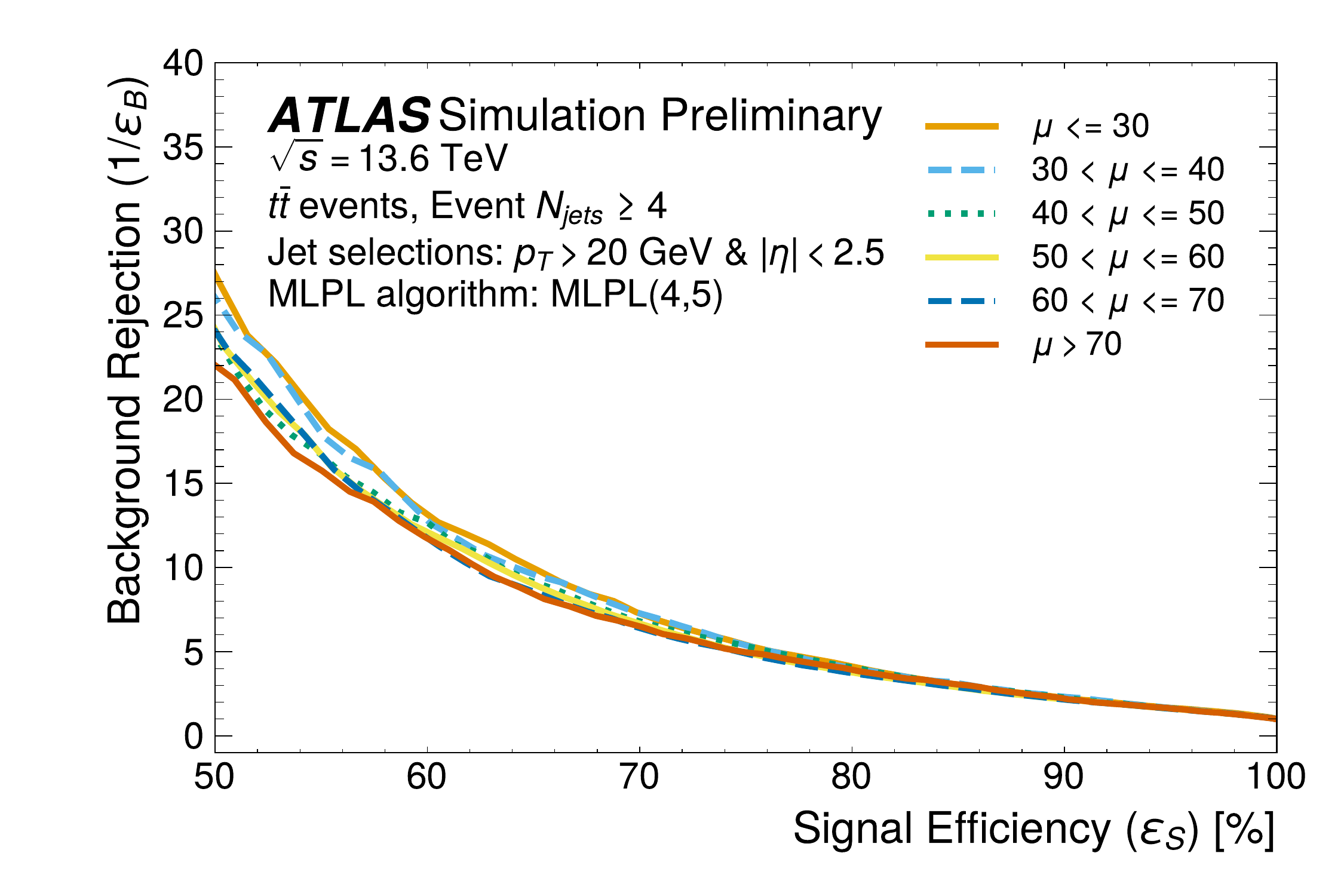}
    \caption{ROC curves showing the performance of the MLPL(4,5) discriminant for a 
    four-jet selection, in bins of the number of interactions per bunch crossing, $\mu$, 
    used as a measure of pile-up. Taken from Ref.~\cite{ATL-DAQ-PUB-2026-002}.}
    \label{fig:roc-curve}
\end{figure}

To evaluate the performance of the MLPL discriminant, a separate sample of simulated semileptonic $t\bar{t}$ events
is used, distinct from all samples involved in the DIPz training and testing. 
Targeting optimisation for a four-jet signature, the signal region is defined as events
containing at least four selected jets, all matched to the same truth collision vertex. The background
region consists of events containing at least four selected jets where no group of four jets is matched
to the same truth collision vertex. The MLPL(4,5) discriminant values are computed for the entire signal
and background datasets (about 282\,000 signal events and about 30\,000 background events). Then,
Receiver Operating Characteristic (ROC) curves are constructed by scanning thresholds on the
discriminant and computing the corresponding signal efficiency and background rejection
(Figure~\ref{fig:roc-curve}), binned in the number of interactions per bunch crossing, $\mu$,
used as a measure of pile-up. The results show a clear separation between signal and background, 
and a considerable background rejection can be achieved, depending on the chosen signal 
efficiency. Moreover, the resulting curves exhibit only a modest dependence on pile-up across the
full efficiency range, indicating that the discriminant maintains similar separation power
from low- to high-$\mu$ conditions. This stability is consistent with the general expectation
that the algorithm targets pile-up-driven contamination while remaining robust against changing
pile-up conditions.

\section{Deployment in the ATLAS HLT}
The DIPz+MLPL approach was implemented in the ATLAS \textsc{Athena} framework~\cite{Athena}
and deployed at the HLT preselection stage of six-jet triggers during Run~3 data taking.
The deployed DIPz was trained with the reconstructed primary-vertex position as a proxy
target, instead of the truth-defined target of Section~2. The truth-defined target is shown 
to improve the resolution and calibration of the DIPz predictions~\cite{ATL-DAQ-PUB-2026-002}.
Online, the discriminant was 
evaluated as MLPL(6,10),
where $m=10$ limits the computational cost in rare events with very high jet multiplicity.
For events with at least six jets with $p_{\mathrm{T}}>20$~GeV, the MLPL discriminant cut lowered 
the background rate by more than a factor of two while losing only 5\% of a representative six-jet benchmark signal,
and it reduced the average CPU execution time of the trigger chain by about 30\%~\cite{ATL-DAQ-PUB-2026-002}.
The rate reduction enables the use of lower jet $p_{\mathrm{T}}$ thresholds while keeping trigger rates within
acceptable limits.

\section{Conclusion}

The use of machine learning, through DIPz, unlocked the task of jet origin regression 
with precise predictions and reliable uncertainty estimates. This enabled the downstream usage 
of a new paradigm in pile-up jet rejection that is agnostic to upstream primary-vertex 
choices, unlike the traditional pile-up mitigation techniques adopted in ATLAS~\cite{JVT}. The 
separation power of the DIPz+MLPL approach is shown to change only modestly across 
the pile-up range of Figure~\ref{fig:roc-curve}, which  motivates studying the 
approach under the HL-LHC conditions, where DIPz+MLPL could be a key handle to fully 
exploit the physics potential of Run 4 data.


\bibliographystyle{unsrt}  
\bibliography{references}  

\end{document}